# Termite-hill: From natural to artificial termites in sensor networks

[1]Adamu Murtala Zungeru, [2]Li-Minn Ang, [3]Kah Phooi Seng

[1]School of Electrical and Electronic Engineering, University of Nottingham,

Jalan Broga, 43500 Semenyih, Selangor Darul Ehsan, Malaysia

[2]School of Engineering, Edith Cowan University,

Joondalup, WA 6027, Australia

[3]School of Computer Technology, Sunway University,

5 Jalan Universiti, Bandar Sunway, 46150 Petaling Jaya, Selangor, Malaysia

**ABSTRACT**

Termites present a very good natural metaphor to evolutionary computation. While each individual's computational power is small compared to more evolved species, it is the power of their colonies that inspires communication engineers. This paper presents a study of artificial termites in sensor networks for the purpose of solving its' routing problem. The behaviours of each of the termites in their colony allow their simulation in a restricted environment. The simulating behaviour demonstrates how the termites make use of an autocatalytic behaviour in order to collectively find a solution for a posed problem in reasonable time. The derived algorithm termed Termite-hill demonstrates the principle of termites' behavior to routing problem solving in the real applications of sensor networks. The performance of the algorithm was tested on static and dynamic sink scenarios. The results as compared with other routing algorithms and with varying network density show that Termite-hill is scalable and improved on network energy consumption with a control over best-effort-service.

**Keywords**- Swarm Intelligence, Wireless Sensor Network, Termite Colony Optimization, Termite-hill, Artificial Termites, Simulation, Target Tracking

## 1. INTRODUCTION

Termites are relatively simple beings. With their small size and small number of neurons, they are incapable of dealing with complex tasks individually. The termite colony on the other hand is often seen as an intelligent entity for its great level of self-organization and the complexity of tasks it performs. In this paper, we will focus on one of the resources termite colonies use for their achievements, pheromone trails, and furthermore, show the similarity between termite colonies and sensor networks. We also try to show some relationship between the stigmergic behaviour facilitated by pheromones and the process of representation in a complex system (sensor network). One way termites communicate is by secreting chemical agents that will be recognized by receptors on the bodies of other termites. For example, a termite is capable of determining if another termite is a member of its own colony by the "smell" of its body. One of the most important of such chemical agents is the pheromone. Pheromones are molecules released from glands on the termite's body. Once deposited on the ground they start to evaporate, releasing molecules of that chemical agent into the air. Individual termites leave a trail of such scent, which stimulates other termites to follow that trail, dropping pheromones while doing so (Matthews & Mattheus, 1942). This use of the environment as a medium for indirect communication is called *stigmergy*. This process will continue until a trail from the termite colony to the food source is established. The creation of a trail with the shortest distance from nest to food source is a side effect of their behaviour, which is not something they have as an *a priori* goal. While following very basic instincts, termites accomplish complex tasks for their colonies in a perfect demonstration of emergent behaviour. In the foraging example, one of the characteristics of the pheromone trail is that it is highly optimized, tending toward the shortest highway between the food source and the termites' nest (hill).

However, a sensor network is an infrastructure composed of sensing, computing and communication elements that give a user or administrator the ability to instrument, observe and react to events and phenomena in a specific environment (Saleem et al., 2010; Zungeru et al., 2012b; Akyildiz et al., 2002). Wireless Sensor Networks (WSNs) are collections of compact-size, relatively inexpensive computational nodes that measure local environmental conditions, or other parameters and forward such information to a central point for appropriate processing using radio frequency (RF) transceivers attached to them. Each sensor node is equipped with embedded processors, sensor devices, storage devices and radio transceivers. Nevertheless, the sensor nodes typically have limited resources in terms of battery supplied energy, processing capability, communication bandwidth, and storage. WSN nodes can sense the environment, communicate with neighboring nodes and in many cases perform basic computations on the data being collected. WSNs applications range from commercial applications such as healthcare, target tracking, monitoring, smart homes, surveillance applications and intrusion detection. The main problem in WSN is how to design a routing protocol which is not only energy efficient, scalable, robust and adaptable, but also provides the same or better performance than that of the existing state-of-the-art routing protocols.

Social insect communities have many desirable properties from the WSN perspective as surveyed in (Zungeru et al., 2012b; Saleem et al., 2010). These communities are formed from simple, autonomous, and cooperative organisms that are interdependent for their survival. Despite a lack of centralized planning or any obvious organizational structure, social insect communities are able to effectively coordinate themselves to achieve global objectives. The behaviors which accomplish these tasks are emergent from much simpler behaviors or rules that the individuals are following. The coordination of behaviors is also adaptive, flexible and robust, and necessary in an unpredictable world which is capable of solving real world problems. The complexity of the solutions generated by such simple individual behaviors indicates that the whole is truly greater than the sum of the parts (Roth & Wicker, 2003; Hölldobler & Wilson, 1990). The characteristics described above are desirable in the context of sensor networks. Such systems may be composed of simple nodes working together to deliver messages, while resilient against changes in its environment. The environment of sensor network might include anything from its own topology to physical layer effects on the communications links, to traffic patterns across the network. A noted difference between biological and engineered networks is that the former have an evolutionary incentive to cooperate, while engineered networks may require alternative solutions to force nodes to cooperate (Buttyan & Hubaux, 2000; Mackenzie & Wicker, 2001). In general, such self organization of biological species is known as swarm intelligence. Research on this field of swarm intelligence has been focused on working principles of ant colonies as adopted in (Bonabeau et al., 1999; Dorigo & Di Caro, 1998), slime mold (Li et al., 2011) and honey bees (Saleem & Farooq, 2005). To the best of our knowledge, little attention has been paid in utilizing the organization and behavioral principles of other swarms such as termites to solve real world problems. In this approach, termite agents were modeled to suit the energy resource constraints in WSNs for the purpose of finding the best paths between sites as a function of the number of visited nodes and the energy of the path, by extensively borrowing from the principles behind the termite communication.

Since communication is an energy expensive function, given a network and a source-destination pair, the problem is to route a packet from the source to the destination node using minimum number of nodes, low energy, and limited memory space so as to save energy. It then implies that when designing a routing protocol for WSN, it is important to consider the path length as well as energy of the path along which the packet is to traverse before its arrival at the sink, while also maintaining low memory usage at the network nodes. In Termite-hill, termite agents are considered as packets that travel the network changing routing information in order to find the best path towards the termite-hill, in this case towards the sink node. The hill is a specialized node called sink node. In this work, which is a continuation of our earlier work, we will show that the Termite-hill routing algorithm is scalable, robust, adaptable and above all energy efficient with less latency.

The rest of the paper is organized as follows. Section 3 presents a brief description of simulating the behaviors' of termites. In Section 4, we describe Termite-hill routing algorithm. Section 5 evaluates the performance of Termite-hill and other routing protocols. Section 6 concludes the paper with comments for future work.

## 2. RELATED WORK

The idea of using a swarm paradigm to establish routes in communication networks is not new. The artificial intelligence community is seeing a shift toward techniques based on evolutionary computation. Inspiration comes from several natural fields such as genetics, metallurgy (simulated annealing) and the mammal immune system. Growing interest in ant colony and swarm algorithms is further demonstration of this new trend.

Marco Dorigo leads the research on optimization techniques using artificial ant colonies (Dorigo et al., 1999). Since 1998, Dorigo has been organizing a biannual workshop on Ant Colony Optimization and swarm algorithms at the *Université Libre de Bruxelles*. Dorigo and his colleagues have successfully applied ant algorithms to the solutions of difficult combinatorial problems such as the travelling salesman problem, the job scheduling problem and others. In (Ramos & Almeida, 2000) and Semet et al. (2004), ant colony approach is used to perform image segmentation, also in a related work, Heusse et al. (1998) and Merloti (2004), applied concepts of ant colonies on routing of network packages.

In simulation, ant colony behavior offers clear demonstration of the notion of emergence with complex system of which coordinated behavior can arise from the local interactions of many relatively simple agents. Stigmergy appears to the viewer almost intentional, as if it were a representation of aspects of a situation. Yet, the individuals creating this phenomenon have no awareness of the larger process in which they participate. This is typical of self-organizing properties: visible at one level of the system and not at another. Considering this, Lawson & Lewis (2004) have suggested that representation emerges from the behavioral coupling of emergent processes with their environments. We hope here to reveal, through experiments with a simple termite colony, the variety of parameters which affect this self-organizing tendency.

In Sensor driven and Cost-aware ant routing (SC) (Zhang et al., 2004), it is assumed that ants have sensors so that they can smell where there is food at the beginning of the routing process so as to increase possibility of sensing the best direction that the ant will go initially. In addition to the sensing ability, each node stores the probability distribution and the estimates of the cost of destination from each of its neighbors. But it suffers from misleading when there is obstacle which might cause errors in sensing. In their extended work, Flooded Forward ant routing (FF), Zhang et al. (2004) argued the fact that ants even augmented with sensors, can be misguided due to the obstacles or moving destinations. The protocol is based on flooding of ants from source node to the sink node. In the case where destination is not known at the beginning by the ants, or cost cannot be estimated, the protocol simply use the broadcast method of sensor networks so as to route packets to the destination. Probabilities are updated in the same way as the basic ant routing, though, FF reduces the flooding ants when a shorter part is transverse. However, the authors only focused on the building of an initial pheromone distribution, which is good at system start-up, but bad when the system density is high. Among other protocols used for comparison purpose is a popular classical routing protocol, Ad-hoc On-demand Distance Vector (AODV) (Charles et al. 1999). Furthermore, our focus is on the routing packets problem, while in their work the authors focused on the optimal movement of mobile sensors.

Besides all the drawback of each of the related protocols, almost all the algorithms tend to scarify the network performance as against the improvement of energy consumption of the nodes, and vice-versa for others with less scalability and adaptability.

## 3. SIMULATING THE BEHAVIOR OF TERMITES

Simulation according to (Wikipedia, 2003) is the imitation representation of the functioning of one system or process by means of the functioning of another. Many computer simulations try to imitate some real-world systems or processes as accurately as possible. Though, in many cases, computer simulations are used to make predictions about real-world processes. In this section, we program artificial termites so as to investigate their termite-like behaviors. The main target is to simulate the termite world, and this will probe some challenges that will be helpful in solving the routing problem in wireless sensor networks. Thus with the increasing interest for social insects, lots of people tend to be fascinated with the general behavior of ants. Thus an ant as an individual performs a simple act, but with the collection of many (colony), they perform rather sophisticated behavior. Thus termites as a subset of ants have come to be viewed as a prototypical example of how complex group behavior can arise from simple individual behavior. As such, the relationship between the colony and termites can be seen as an illuminating model, or at least an inspiring metaphor for thinking about other group or individual relationships, such as the relationship between an organ and its cells, a cell and its macromolecules, a corporation and its employees, or a country and its citizens (Resnick, 1994, 1997).

Again, each termite colony has a queen, unlike the ant system whereby the queen serves as leader. The termite queen does not give directives to the workers as in ants. Though, "Queen" seems to imply "Leader", but it is more of mother than a leader to the colony. Also, ant colony algorithms are mainly designed having in mind the proactive nature of the colony, while this termite's algorithm is reactive in nature. Detail explanation of this relationship is given in Section 4, and interested readers on the similarities and differences between ant and termite algorithm can refer to Zungeru et al. (2012b). It is worth knowing that on the termite hill building site, the termite has no site engineer (leader), that is to say that there is no one to take control of the master plan. Even with that, each of the individual termites carries out a specific simple task. Being practically blind and they must interact, they does that through their senses of smell and touch. Through their local interactions, an important feature opens up. The principle of hill building, through cooperative behavior without site engineer to give directives, makes them suited for solving routing problem in sensor network where information is expected to be gathered in one place (sink). This means that simulating the construction of an entire termite nest will give more insight on their behavior, and thus can easily be mapped to simulating the sensor network. As such, in this section, we program some artificial termites to collect wood samples and the wood samples are expected to be gathered into particular sites (hills). Though, real termites do not actually carry wood samples from place to place, rather, they eat pieces of woods, then build hills with the feces they produce from the digested wood. The main challenge which is the motivating factor in this work is how to figure out a decentralized strategy for adding some order to a disordered collection of wood samples. Initially, the wood samples are randomly distributed throughout the termites' environment, but as the program runs, the termites are expected to organize the wood samples into a few orderly piles. With this model, we could map this to sensor network of which sensor nodes are distributed haphazardly with the aim to sense their environment and to gather the sensed event into one place (sink). Following the four

(4) rules bounded by each termite as proposed in (Zungeru et al., 2012a), we then program the termites to gather the disordered wood samples into an ordered form, and into fewer piles. In this program of termites piling up the disordered wood samples into order, we wrote set of programs for different functions. This set of functions include: (1) defining sets of variables and initializing the global variables and functions. This includes the number of woods needed in the termites' world to the dimension of the termites' environment. In this, the number of woods equals the number of potential hills in the environment. (2) A function to distribute the wood and termites in the simulation environment. (3) Function definitions. (4) Function to make the termites move in the simulation environment, and (5) function to make termites pick up and put down the woods in a piles. As the termites pick up woods and look for piles, they do so in an orderly manner in which they put down wood samples only at a place where there exist at least a sample of wood. That is to say that, they do not put down the wood samples in an empty space. This process leads to the gathering of wood into fewer piles. If all of the wood samples from a particular pile are by chance removed completely due to the fact that all the wood samples are completely removed from that point, it then means that, termites will never drop any wood sample in that spot. This means that, that particular hill will not by chance grow again. If there happen to be an existence of a pile or hill, its size will have the probability of increasing or decreasing, though, the existence of a pile once gone, it is gone forever. With this behavior, termites are able to gather the disordered wood into ordered forms. As an example of this behavior, we simulate 200 termites and 100 wood samples in a 200m by 200m DMZ (De-Militarized Zone) application environment, and we further increase the number of termites up to 500 in the environment. In Figure 1, we show the results gotten from the simulation. The graph shows clearly their behaviors with respect to simulation time and number of termites in the environment to gather the widely dispersed woods. As described above, the following pseudo-code (Code 1.0) explains the process of the program.

**Code 1.0:** Simulation of artificial termites in a real-world behavior Algorithm

```
1.  //Termite's real world behaviour:
2.  //Define variable and Initialization
3.   int Termite=200;
4.  int woodchip=100;
5.  pile-name;
6.  Pile-wood-count;       //indicates the number of woods each pile is constructed with.
7.  int Number-of-pile ;
8.  Int x=200, y=200;      // the environment in which termite and wood chips are distributed.
9.   //functions' prototype
10. Distribute_wood (wood_chip,x,y);
```

```
11. View_wood ();
12. Distribute_termite (Termite,Distribute_wood,x,y);
13. Termite-move();
14. //main
15. Void main (){
16.   Distribute_wood;
17.   Distribute_Termite;
18.   //call Distribute_wood function for distributing woods randomly
19.   Distribute_wood=Distribute_wood (wood_chip,x,y);
20.   //Distribute_ termite function is in charge of distributing termite and pick up and put
21.   down the woods
22.   Distribute_termite=Distribute_termite (Termite,Distribute_wood,x,y);
23.   Print ("number of pile = " number-of-pile);
24.   Print (Pile$i,Pile$i.pile-wood-count)
25.   Wood-in-piles=0
26.   For (c=0;c<number-of-pile,c++){
27.         Int Wood-in-piles =Wood-in-piles +Pile$c.pile-wood-count
28.   }
29.   Print Wood-in-piles;
30. }
31. // Functions' definition:
32. Function Distribute_wood (wood_chip,x,y){
33.   int number-of-wood=0;
34.   for(int i = 1; i <= woodchip; i++)
35.   {
36.         L1:     int wood-x ← choose random number between 0 to x;
37.                 int wood-y ← choose random number between 0 to y;
38.                 check (x,y) ;
39.                 if the place is empty{
40.                    put wood there
41.                    number-of-wood=number-of-wood++;
42.                    Number-of-pile =  number-of-wood;
43.                    /*in the initialization step, each wood determines a pile, and therefore,
44.                    when we find an empty place for wood,
45.                    we should keep the coordinate in the array for storing the
46.                    pile's location*/
47.                    matrix[wood-x][wood-y]="Pile$i";
48.                    pile-name[i] = "Pile$i";
49.                    pile-wood-count[$i] = 1;
50.                 }
51.                 else
52.                    goto L1;
53.   }
54.   Return   number-of-wood;
55. }
56. Function Distribute_termite (Termite,Distribute_wood,x,y){
57.   int  pick_up_wood = 0 ; //indicate how many woods are been carried by termites
58.   // distributes the termite in the environment
59.   For(i= 0 ; i<Termite;i++){
60.         L2: int Termite-x ← choose random number between 0 to x;
61.             int Termite-y ← choose random number between 0 to y;
62.             check (x,y) ;
63.             if the place is empty Put termite in (x,y)
64.             else goto L2 ;
65.   }
66. While (simulation's time > 0){
67.   //Termites should keep moving until they find a wood
```

```
68.    Termite-move();
69.    // termite find a wood
70.    pick_up_wood = pick_up_wood + 1;
71.    Pile$i.pile-wood-count=Pile$i.pile-wood-count - 1;
72.    If (Pile$i.pile-wood-count < 1){
73.       Delete Pile$i;
74.       Number-of-pile = Number-of-pile-1;
75.    }
76. // termite should keep a random movement until they find another wood and put down this
77.    one near it.
78.    Termite-move();
79.    Select the nearest empty place
80.    Put the wood;
81.    Pile$i.pile-wood-count=pile-wood-count+1;
82.    Termite-move();
83. }
84. Function Termite-move(){
85. Int row = Termite-x;
86. Int col  = Termite-y;
87. L4:
88.        For (row; row<x;row++){
89.        For (col;col<y;col++){
90.           If ((row,col ) == (wood-x,wood-y)){
91.               Return (Pile$i);
92.                Break;
93.           }
94.        }
95.     }
96. // that means termite did not find wood and it reaches (200,200)
97. Col=0;
98. Row=0;
99. Goto L4;
100.}
```

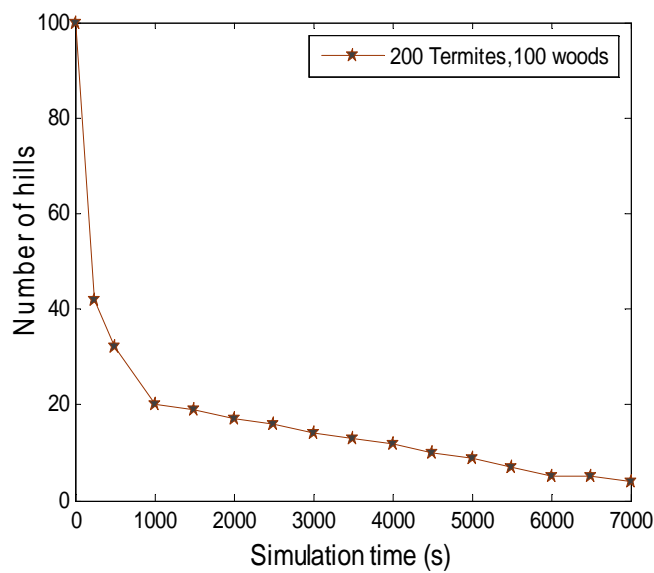

(a)

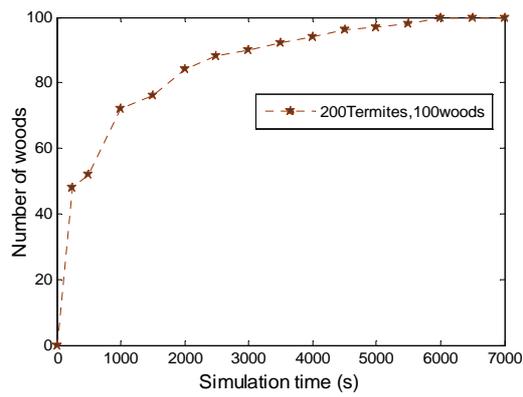

(b)

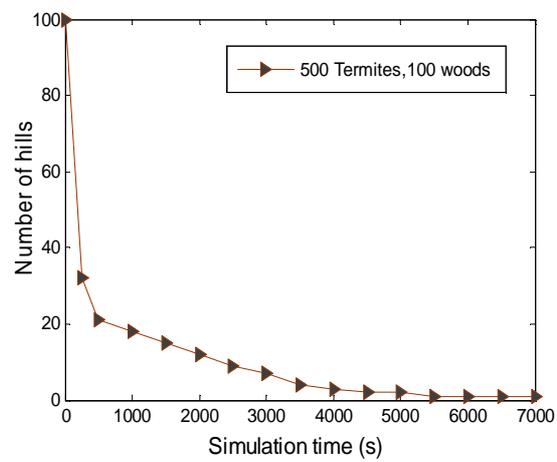

(c)

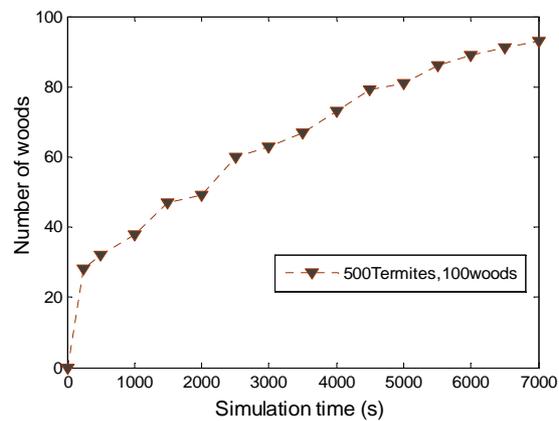

(d)

Figure 1. Behavioral pattern of termites gathering wood samples in the presence of variation in simulation time with respect to: (a) fewer termites and number of hills built (b) fewer termites and number of woods gathered (c) more termites and number of hills built and (d) more termites and number of woods gathered.

At the initial runs of the program when we run the simulation, termites put down the wood samples next to each other wood samples, rather than on top of other samples in the normal hill building. But to us, it is still fine as long as the wood samples are gathered into one or fewer piles. Though, piles are not clearly defined as termites pick up wood samples from the middle of well established piles such that what are remained could be seen as two or more piles with one pile of higher concentration of woods. Though, as we keep running the simulation, initially, the wood samples are gathered into tenths of samples, but as the simulation proceeds, the number of samples per pile increases while the number of piles in the environment decreases. This action can be seen in Figure 1(a-d). It was also observed that, after 1000 seconds of simulation time with 200 termites randomly distributed along with 100 wood samples in the environment, there were about 20 piles of woods out of the initial piles of 100, with a total of 72 woods. After 5000 seconds, 9 piles were recorded with a total of 97 woods, and this continues, and as the simulation time proceeds to 7000 seconds, 4 piles were recorded and a total of 100 woods were also recorded. However, it then means that with more time of simulation, the number of piles shrink to just a single whole pile. Also, the number of woods gathered (success rate) tends to 100%. This is shown in Figure 1(a) and (b). But the shrinking of the piles as observed with fewer termites of 200 was not a fast one as expected. We then increased the number of termites to 500 as against the original 200 in the first case. In this, after 1000 seconds, we recorded 18 piles with a total of 38 woods, and after 5000 seconds, we recorded 2 piles with a total of 81 woods, and also at 7000 seconds, a large pile of 1 was recorded with 92 woods, which implies that 7 woods were still carried by some termites. This is expected since in most cases, with a high population of termites in the environment, the piles shrink faster which means that the latency reduces and the success rate diminishes along. In all, the ability of termites to gather woods into fewer piles is the convergence of the network when we have more termites in the environment. But there is a threshold for the number of termites as observed in the experiment for congestion control and to avoid the low success rate level. It was also observed in the experiment that as we increase the number of termites above 5 times the number of wood samples, the environment gets congested and all woods are carried by the termites of which it becomes difficult for them to form any reasonable pile. With these observations in their behavior, we also testify the assumption made in (Zungeru et al., 2012a) in the reduction of ants in the network for congestion control. With this behavior and observations, we then map our findings into sensor network which will be described in Section 4.

## 4. THE TERMITE-HILL ROUTING ALGORITHM

Termite-hill is a routing algorithm for wireless sensor networks (WSNs) that is inspired by the termite behaviors. Preliminary results of this algorithm are reported in (Zungeru et al., 2012a). Analogous to the termite ad-hoc networking (Roth and Wicker, 2003), each node serves as router and source, and the hill is a specialized node called sink which can be one or more depending on the network size. Depending on the network, each network node can also serve as a termite hill. Termite-hill discovers routes only when they are required. When a node has some events or data to be relayed to a sink node and it does not have the valid routing table entry, it generates a *forward soldier* and broadcasts it to all its neighbors. When an intermediate node receives this *forward soldier*, it searches its local routing table for a valid route to the requested destination. If the search is successful, the receiving node then generates a *backward soldier* packet, which is then sent as a unicast message back to the source node where the original request was originated using the reverse links. If the node has no valid route to the destination, it sets up a reverse link to the node from which the *forward soldier* was received and further broadcasts the *forward soldier* packet. When the destination node receives the *forward soldier* packet, it generates a *backward soldier* packet which is also unicast back to the source node. On reception of the backward *soldier* packet, each intermediate node updates its routing table to set up a forward pointer and relays the *backward soldier* message to the next hop using the reverse pointer. The process continues till the *backward soldier* is received by the original source node. For Termite-hill algorithm for WSNs, *HELLO* packets are not used to detect link failures. Rather it uses feedback from the link layer (MAC) to achieve the same objective. Intermediate nodes do not generate *backward soldier* even if they have a valid route which avoids the overhead of multiple replies. It also employs cross layer techniques to avoid paths which have high packet loss. As such the termite-hill is designed to function in three modules. In the course of the algorithm design, the following assumptions were also made: 1. each node is linked to one or more nodes in the network (neighbors), 2. A node may act as a source, a destination, or a router for a communication between different pair of nodes, 3. Neither network configuration nor adjacency information is known before hand, and 4. The same amount of power is required for sending a message between any pair of adjacent nodes throughout the network.

4.1 The Pheromone Table

The pheromone table keeps the information gathered by the forward soldier. Each node maintains a table keeping the amount of pheromone on each neighbor path. The node has a distinct pheromone scent, and the table is in the form of a matrix with destination nodes listed along the side and neighbor nodes listed across the top. Rows correspond to destinations and columns to neighbors. An entry in the pheromone table is referenced by $T_{n,d}$ where *n* is the neighbor index and *d* denotes the destination index. The values in the pheromone table are used to calculate the selecting probabilities of each neighbor. From Figure 2 below, when a packet arrives at node *G* from previous hop *S*, i.e. the source, the source pheromone decay, and pheromone is added to link $\overrightarrow{SG}$. Backward soldier on their way back from the sink node is more likely to take through *G*, since it is the shorter path to the destination i.e. $\overrightarrow{SGED}$. The pheromone table of node *G* is shown in Figure 2 below with nodes A, S, F, and E as its neighbor. It is worth noting that all neighbors are potential destinations. At node G, the total probability of selecting links $\overrightarrow{ED}, \overrightarrow{FE}, \overrightarrow{AC}$ or $\overrightarrow{SB}$ to the destination node is equal to unity (1) i.e. $\sum T_{ED} + T_{SD} + T_{AD} + T_{FD} = 1$. It will then be observed that, since link $\overrightarrow{GED}$ is shorter to the destination for a packet at node G, more pheromone will be present on it and hence, soldiers are more likely to take that path.

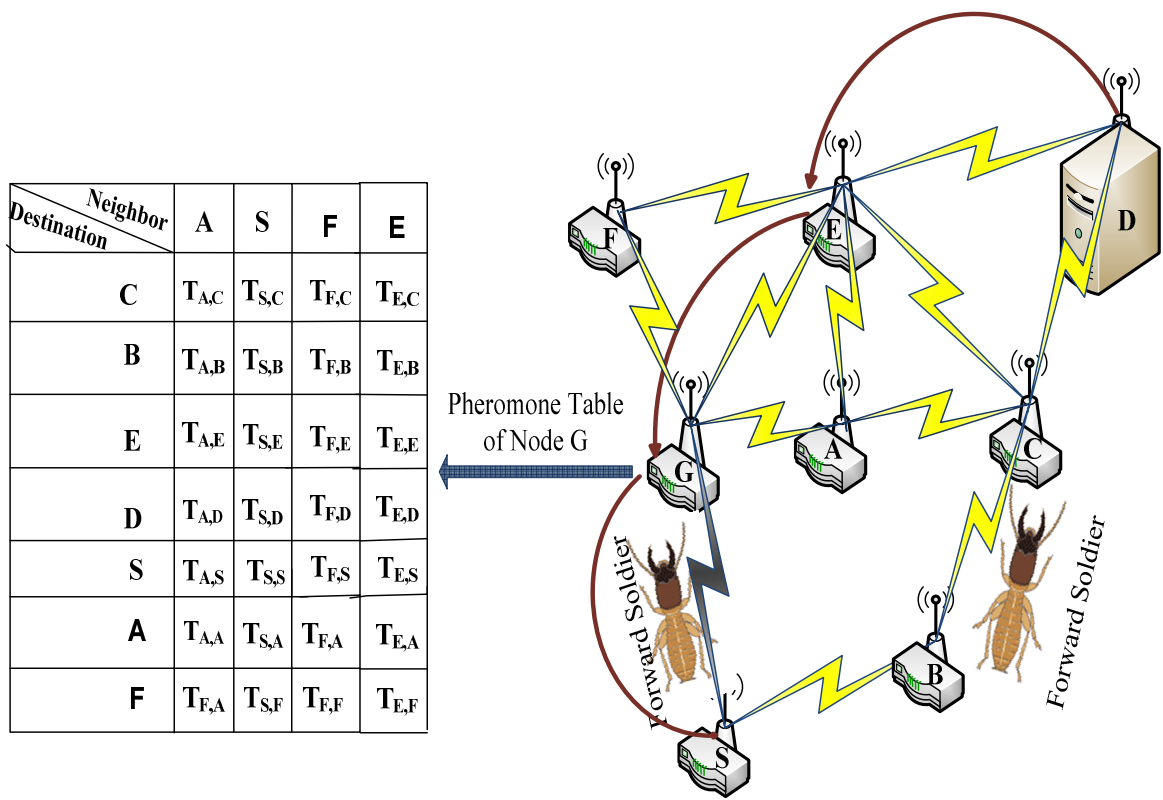

**Figure 2.** Description of pheromone table of node *G*

4.1.1 Pheromone Update

When a packet arrives at a node, the pheromone for the source of the packet is incremented by $\gamma$, where $\gamma$ is the reward. Only packets addressed to a node will be processed. A node is said to be addressed if it is the intended next hop recipient of the packet. Equation (1) describes the pheromone update procedure when a packet from source *s* is delivered from previous hop *r*. A prime indicates the updated value.

$$T'_{r,s} = T_{r,s} + \gamma \tag{1}$$

Where

$$\gamma = \frac{N}{E - \left(\frac{E_{min} - N_j}{E_{av} - N_j}\right)} \tag{2}$$

Where $E$ is the initial energy of the nodes, $E_{min}$, $E_{av}$ are the minimum and average energy respectively of the path traversed by the forward soldier as it moves towards the hill. The values of $E_{min}$ and $E_{av}$, depends on the number of nodes on the path and the energy consumed by the nodes on the path during transmission and reception of packets. The minimum energy of the path ($E_{min}$) can be less than the number of nodes visited by the forward soldier, but the average energy of the path ($E_{av}$) can never be less than the number of visited nodes. $N_j$, represents the number of nodes that the forward soldier has visited, and $N$ is the total number of network nodes.

4.1.2 Pheromone Evaporation

Pheromone is evaporated so as to build a good solution in the network. Each value in the pheromone table is periodically multiplied by the evaporation factor $e^{-\rho}$. The evaporation rate is $\rho \geq 0$. A high evaporation rate will quickly reduce the amount of remaining pheromone, while a low value will degrade the pheromone slowly. The nominal pheromone evaporation interval is one second; this is called the decay period. Equation (3) describes the pheromone decay.

$$T'_{n,d} = T'_{n,d} * e^{-\rho} \tag{3}$$

Though for robustness and flexibility some application needs a slow decay rate, and some applications like security and target tracking applications, need fast decay process and will determine the value of the decay period. That is, the value of $\rho$ and $x$ in equation (4) depends on the application area. Hence to account for the pheromone decay each value in the pheromone table is periodically subtracted by percentage of the original value as shown in equation (4).

$$T'_{n,d} = (1 - x)T'_{n,d} \tag{4}$$

Where, $0 \leq x \leq 1$

If all of the pheromone for a particular node decays, then the corresponding row and/or column are removed from the pheromone table. Removal of an entry from the pheromone table indicates that no packet has been received from that node for quite some time. It has likely become irrelevant and no route information must be maintained. A column (destination listing) is considered decayed if all of the pheromone in that column is equal to a minimum value. If that particular destination is also a neighbor then it cannot be removed unless all entries in the neighbor row are also decayed. A row is considered decayed if all of the pheromone values on the row are equal to the pheromone floor. Neighbor nodes must be specially handled because they can forward packets as well as originate packets. A decayed column indicates that no traffic has been seen which was sourced by that node. Since neighbors can also forward traffic, their role as traffic sources may be secondary to their role as traffic relays. Thus, the neighbor row must be declared decayed before the neighbor node can be removed from the pheromone table. If a neighbor is determined to be lost by means of communications failure (the neighbor has left communications range), the neighbor row is simply removed from the pheromone table. Following the pheromone update and evaporation, there is a pheromone limits which are the *pheromone ceiling*, the *pheromone floor*, and the *initial pheromone*.

4.2 Route Selection

Each of the routing tables of the nodes is initialized with a uniform probability distribution given as;

$$P_{s,d} = \frac{1}{N} \tag{5}$$

Where $P_{s,d}$ is the initial probability of each source node, and it represents the probability by which an agent at source node *s* take to get to node *d* (destination), and *N* is the total number of nodes in the network.

The equation below details the transformation of pheromone for *d* on link *s* $T_{s,d}$ into the probability $P_{s,d}$ that the packet will be forwarded to *d*.

$$P_{s,d} = \frac{(T_{s,d}+\alpha)^\beta}{\sum_{i=1}^{N}(T_{i,d}+\alpha)^\beta} \tag{6}$$

As pointed out in Figure 2, the summation of the probabilities of taking all parts leading to the destination node is unity (1). The parameters $\alpha$ and $\beta$ are used to fine tune the routing behavior of Termite-hill. The value of $\alpha$ determines the sensitivity of the probability calculations to small amounts of pheromone, $\alpha \geq 0$ and the real value of $\alpha$ is zero. Similarly, $0 \leq \beta \leq 2$ is used to modulate the differences between pheromone amounts, and the real value of $\beta$ is two. But for each of the *N* entries in the node *k* routing table, it will be $N_k$ (where $N_k$ represents neighboring nodes of node *k*) values of $P_{s,d}$ subject to the condition:

$$\sum_{s \in N_k} P_{s,d} = 1; \quad d = 1, \dots, N \tag{7}$$

## 4.3 Termite-hill Agent Model and Modules design

The termites evaluate the quality of each discovered path to a hill by the pheromone contents of the pebbles on the path. This means that, not all the discovered path receives reinforcement. *Termite-hills* works with three types of agents: reproductive, soldiers and workers. The algorithm is designed to function as three main modules*: route discovery, seed, and data.* Below is the pseudocode describing the operation of the algorithm, and it is divided into four parts as shown in Pseudo-code 2 to 5.

| **Code 2:** Route Discovery Pseudocode |
|---|

```
1.  Required:   A copy of Forward Soldier (FS)
2.  if    (SinkNode)      then
3.         // Upload Payload and pass to application layer
4.         PayloadToApplication (FS);
5.         UpdateForwardingTable (FS.From, NodeID, PathID);
6.         // Construct a Backward Soldier and forward to FS.From
7.         BS ← ConstructBackwardSoldier (BS);
8.         Forward (BS, FS.From);
9.  else if    (NotSeenBefore (FS))    then
10.        N_j ← FS.Hops ← FS.Hops + 1;
11.        if    (FS.Hops ≤ H_max) then
12.            // Set Broadcast Flag
13.            BFlag ← 1;
14.        else
15.            BFlag ← StochasticForwarding ( );
16.        end if
17.        N ← TotalNetworkNodes ← (Node.Total);
18.        E ← InitialNodesEnergy ← (Node.Energy);
19.        E_min ← FS.MinEnergy ← Min (FS.MinEnergy, Node.Energy.Min);
20.        E_av ← FS.AvEnergy ← Av (FS.AvEnergy, Node.Energy.Av);
21.        β ← N / (E - (E_min - N_j)/(E_av - N_j));
22.        UpdateSoldierCache (FS.From, FS.SourceID, FS.SoldierID, BFlag, β);
23.        if    (BFlag) then
24.            Broadcast (FS);
25.        else
26.            DeleteForwardSoldier (FS);
27.        end if
28.  else
29.        if    ( Forwarded(FS) )    then
30.            N_j ← FS.Hops ← FS.Hops + 1;
31.            E_min ← Min (FS.MinEnergy, Node.Energy.Min);
32.            E_av ← Av (FS.AvEnergy, Node.Energy.Av);
33.            β ← N / (E - (E_min - N_j)/(E_av - N_j));
34.            if    (β > RewardInSoldierCache ( ))    then
35.                UpdateSoldierCache (FS.From, FS.SourceID, FS.SoldierID, BFlag, β);
36.            end if
37.        end if
38.        DeleteForwardSoldier (FS);
39.  end if
```

**Code 3:** Route Update Pseudocode

1. Required:    A copy of Backward Soldier (BS)
2. **if**   (SourceNode)   **then**
3.     $T_{rs}$ ← CalculatePheromoneValue (BS.Pheromone);
4.     // Update the pheromone and probability tables
5.     UpdatePheromoneTable (BS.From, BS.SinkID, BS.PathID, $T_{rs}$);
6.     UpdateProbabilityTable ($P_{sd}$);
7.     DeleteBackwardSoldier (BS);
8.     // announce path to the neighbors
9.     BroadcastBeacon ( );
10. **else**
11.     // Check for matching BS if earlier forwarded
12.     **if**   (MatchInSoldierCache(BS))   **then**
13.         **//** Update the forwarding table
14.         UpdateForwardingTable (BS.From, SoldierCache, PrevHop, BS.PathID);
15.         Forward (BS, SoldierCache.PrevHop);
16.         DeleteSoldierCacheEntry (BS);
17.         BS.Pheromone ← (BS.Pheromone, Path.Pheromone);
18.     **else**
19.         DeleteBackwardSoldier (BS);
20.     **end if**
21. **end if**

**Code 4:** Working Group Pseudocode

1. Required: A Phenomenon for Transportation to Sink Node
2. **for all**   Phenomenon received from Application layer   **do**
3.     W = Worker ( );
4.     **if**   (W = = NULL) **then**
5.         **if**   (RouteDiscoveringInProgress( ))   **then**
6.             // Route discovery in progress, wait in cache
7.             StorePayloadInCache (P);
8.         **else**
9.             // Route required, initiate forward soldier
10.             LaunchForwardSoldier (FS);
11.         **end if**
12.     **else**
13.         //Worker found, forward to next hop
14.         Forward (W, NextHop);
15.     **end if**
16. **end for**

**Code 5:** Working Group at intermediate nodes Pseudocode

1. Required:    A Worker
2. **if**   SinkNode( )   **then**
3.     PassToApplication (W.P);
4.     AddToWorkersList (W);
5. **else**
6.     Next ← GetNextHop (W.PathID);
7.     Forward (W, Next);
8. **end if**

## 5. PERFORMANCE EVALUATION

This section evaluates the performance of the routing algorithm (Termite-hill) implemented in Routing Modeling Application Simulation Environment (RMASE) (PARC, 2006; Zhang et al., 2006; Zhang, 2005) which is a framework implemented as an application in the Probabilistic Wireless Network Simulator (PROWLER) (Sztipanovits, 2004). The simulator is written and runs under Matlab, thus providing a fast and easy way to prototype applications and having nice visualization capabilities for the experimental and comparison purpose. The simulation parameters used for this particular experiment are as shown in Table 1 below.

Table 1: Analytical and Simulation Parameters

| Parameters | Values |
|---|---|
| Routing Protocol | SC, FF, AODV, Termite-hill |
| Size of Topology (A) | 100 x 100 |
| Distribution of Nodes | Random distribution |
| Number of Nodes (N) | 100 |
| Maximum number of Retransmission (n) | 3 |
| Transmission Range ( R ) | 35 m |
| Data Traffic | Constant Bit Rate (CBR) |
| Data Rate | 250 kbps |
| Propagation model | Probabilistic |
| Energy consumption | Waspmote-802.15.4 |
| Time of topology change | 2 s |
| Simulation Time | 360s |
| Average Simulation times | 10 |

From several results obtained from our simulation work, we observed the following metrics to evaluate the performance of Termite-hill routing algorithm in WSN.

- *Success rate:* it is a ratio of total number of events received at the destination to the total number of events generated by the nodes in the sensor network. We reported it in percentage (%).
- *Energy consumption***:** It is the total energy consumed by the nodes in the network during the period of the experiment (Joules).
- *Energy utilization efficiency***:** It is a measure of the ratio of total packet delivered at the destination to the total energy consumed by the network's sensor nodes (Kbits/Joules).

## 5.1 Performance of Termite-hill with Static Sink

First, we evaluate the performance of Termite-hill and compare with other routing protocols with static sink. In this scenario, we assumed that the sink node is fixed at a particular destination. In our simulation, results as reported in Figure 3 show the performance in term of success rate of events generated in the network, energy consumption of nodes at the end of the experiment and energy utilization efficiency of the respective algorithms with varied network density to ascertain the proposed algorithms' scalability. In term of successful packet delivered at the sink node, it was observed that Termite-hill has its maximum success rate when the network nodes was still few in number (9 nodes) corresponding to the value of 96.4%. Though, this value degrades a little with increase in the number of network nodes and it was seen that, when the number of nodes in the network approaches the value of 100, the success rate approaches 80%. But as against our initial results with fewer nodes, AODV performance is better than SC when the network density increases due to increase in number of nodes. That is, at the value of 100 nodes in the network, SC has a success rate of 51% as against 69% of AODV. The poor performance of these two algorithms was due to flooding of route discovery packets each time of the routing process, as most of its data packets do not actually get to sink even when generated by the source nodes. Though, the performance of FF which was designed for high success rate is still below that of Termites hill as can be observed in Figure 3. It was also observed that Termite-hill performance was higher as compared to the entire algorithm under investigation. Even with the high reliability (high success rate), its performance in term of energy consumption was better than other algorithms which in turn, makes it the most energy efficient. Termite-hill algorithm achieves both high packet successful delivery and energy utilization efficiency as compared to SC, FF, and AODV due to some of its important features as, first, the launch of its soldier carrying the first generated event in which most cases it is able to find routes to the destination in the first attempt; second, it makes use of restrictive flooding which results in quick convergence of the algorithm; third, it maintains a small event cache to queue events while route discovery is in progress; fourth, it utilizes a simple packet switching model in which intermediate nodes do not perform complex routing table lookup as in others, rather packets are switched using a simple forwarding table at a faster rate; and lastly, the updating rule takes into consideration the paths energy, hence the probability of route selection is also a function of paths remaining energy.

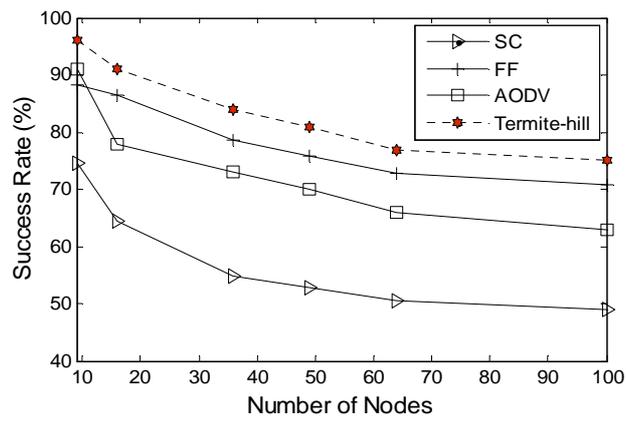

(a)

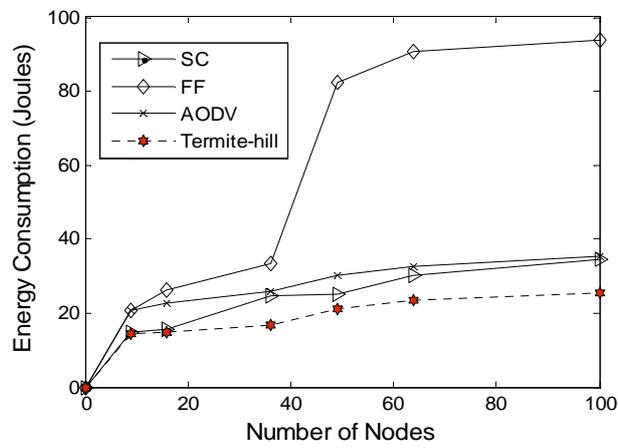

(b)

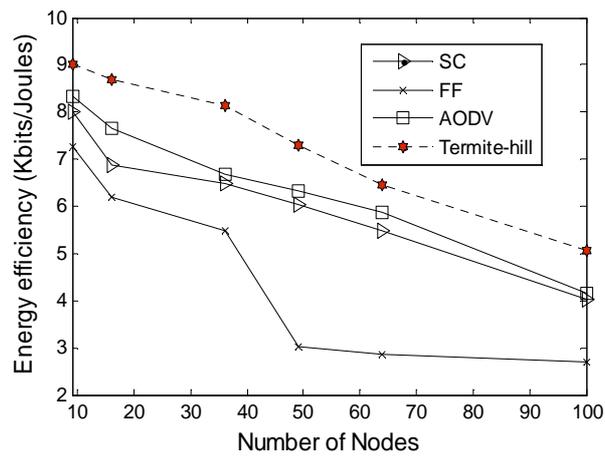

(c)

**Figure 3.** Performance evaluation in static scenario among routing protocols: (a) Success rate (b) Energy consumption (c) Energy Efficiency.

## 5.2     Performance of Termite-hill with a Dynamic Sink (Target Tracking)

In this section, we study and evaluate the performance of Termite-hill with other routing algorithms in dynamic network. In this scenario, we also assumed that the sink can change its location at any given time. The change is not along a path, but in any direction making it different from the mobility scenario. This is basically target tracking scenario. The target in the region of interest has to be monitored, but sometimes, it gets out of transmission range of almost all the nodes, hence the use of dynamic sink became very important as also, sensor nodes needs less hops to get to the sink so as to limit energy consumption. In the first part, we simulate the entire algorithms over long duration of time with fixed speed of sink as shown in Figure 4. In that scenario, Termite-hill performance in terms of successful packet delivery was still higher than the other algorithms. Though, the energy consumption bar of Termite-hill with SC approaches each other with fewer network nodes, but Termites-hill success rate was not comparable to SC even at the fewer nodes in the network. But with its high packet delivery rate, it has the highest energy utilization efficiency as compared to all the algorithms. To further test its performance, we adapt all the routing algorithms in the dynamic scenario with varying network density as shown in Figure 4. In that case, Termite-hill performance in terms of successful packet delivery rate and energy utilization efficiency is higher, with less energy consumption. It will also be observed that though the success rate of each of the routing protocols tends to decrease with increase in network nodes, the energy consumption of all the algorithms also increases as more packets are delivered at the sink node since the average remaining energy keeps on dropping. The poor performance of FF in terms of high energy consumption is due to its pure flooding of Route Request (RREQ) packets (ants), which make it to have unnecessary overhead in the network.

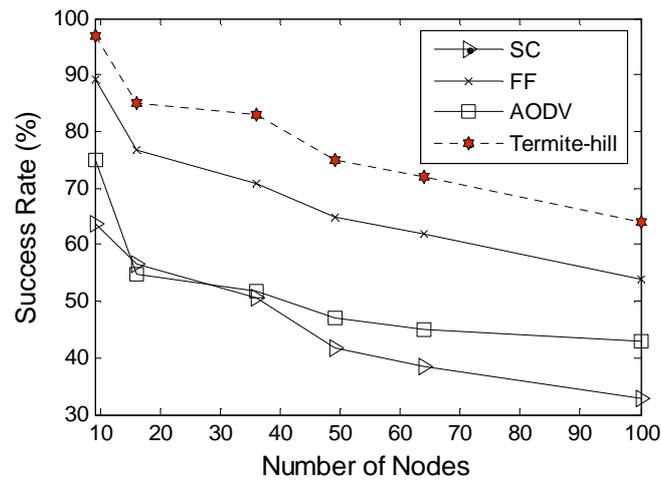

(a)

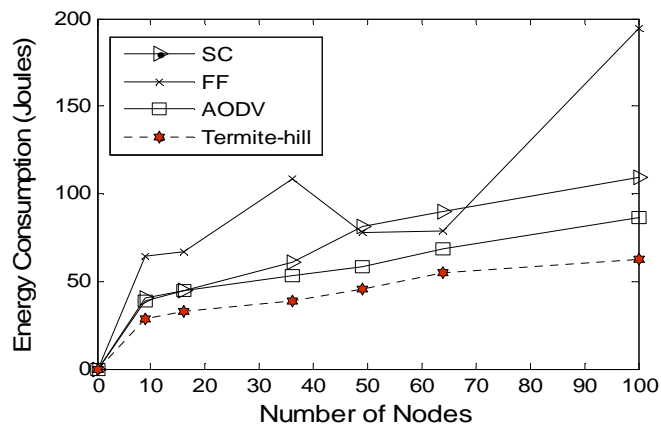

(b)

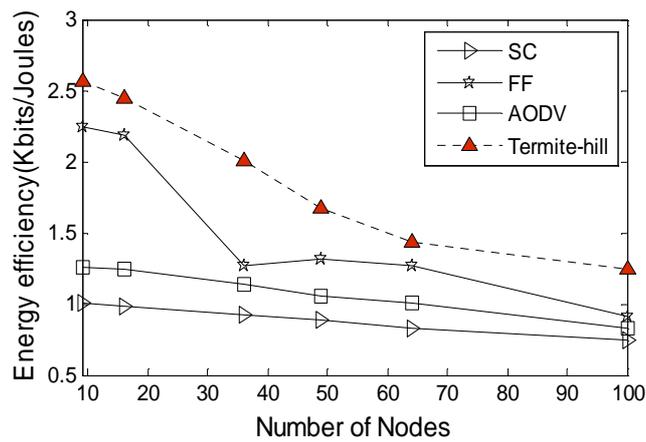

(c)

**Figure 4.** Performance evaluation in Target tracking scenario among routing protocols: (a) Success rate (b) Energy consumption (c) Energy Efficiency

## 6. CONCLUSIONS AND FUTURE WORKS

In this paper, we studied the application of the Termite Colony Optimization metaheuristic to solve the routing problem in wireless sensor networks. A basic Termite based routing protocol was proposed. Several factors and improvements inspired by the features of wireless sensor networks (low energy level, low memory and processing capabilities) were considered and implemented. The resulting routing protocol termed Termite-hill was designed to function in three modules; route discovery, route maintenance and data packet module. The algorithm uses backward and forward soldiers for route discovery and updating between the sensor nodes and the sink node, which are optimized in terms of distance and energy level of each path. The algorithm minimizes network overhead by on-demand routing, and maximizes network reliability and energy savings, which contribute to improving the lifetime of the sensor network. The experimental results showed that the algorithm leads to very good results in different WSN scenarios and the algorithm is overall scalable, robust and above all most energy efficient in comparison with other state-of-the-art routing protocols. We will improve on the Termite-hill routing algorithm based on the readers' comments and suggestions.

# REFERENCES

Akyildiz, I. F., Su, W., Sankarasubramaniam, Y., & Cayirci, E. (2002). Wireless sensor networks: A survey. *Computer Networks*, *38*(4), 393–422. doi:10.1016/S1389-1286(01)00302-4

Bonabeau, E., Dorigo, M., & Theraulaz, G. (1999). *Swarm intelligence: From natural to artificial systems* (pp. 1–278). London, UK: Oxford University Press.

Buttyan, L., & Hubaux, J.-P. (2000). Enforcing service availability in mobile ad-hoc WANs. In *Mobile and Ad Hoc Networking and Computing,* (pp. 87-96). doi:10.1109/MOBHOC.2000.869216

Camilo, T., Carreto, C., Silva, J. S., & Boavida, F. (2006). An energy-efficient ant based routing algorithm for wireless sensor networks. In *Proceedings of 5th International Workshop on Ant Colony Optimization and Swarm Intelligence*, Brussels, Belgium, (pp. 49-59).

Çelik, F., Zengin, A., & Tuncel, S. (2010). A survey on swarm intelligence based routing protocols in wireless sensor networks. *International Journal of the Physical Sciences*, 5(14), 2118-2126.

Dorigo, M., & Di Caro, G. (1998). AntNet: Distributed stigmergetic control for communications networks. *Journal of Artificial Intelligence Research*, *9*, 317–365.

Dorigo, M., Di Caro, G., & Gambardella, L.M. (1999). Ant Algorithms for Discrete Optimization, *Artificial Life*, 5(3), 137-172.

Heusse, S. Guérin, Snyers, D. & Kuntz, P. (1998). Adaptive Agent-driven Routing and Load Balancing. Communication Networks, Technical Report RR-98001-IASC, Department Intelligence Artificielle et Sciences Cognitives, ENST Bretagne.

Hölldobler, B., & Wilson, E.O. (1990). *The Ant,* Harvard University Press, 1990.

Lawson, B.J. & Lewis, J. (2004). Representation Emerges from Coupled Behavior. *Self-Organization, Emergence, and Representation Workshop, Genetic and Evolutionary Computation Conference Proceedings,* Springer-Verlag 2004.

Li, K., Torres, C. E., Thomas, K., Rossi, L. F., & Shen, C. -C. (2011). Slime mold inspired routing protocols for wireless sensor networks. *Swarm Intelligence*, *5*(3-4), 183-223.

MacKenzie, A.B., Wicker, S.B. (2001). Game theory in communications: motivation, explanation, and application to power control. *GLOBECOM'01, IEEE Global Telecommunications Conference,* 2, (pp. 821-826).

Matthews, R.W., & Mattheus, J.R. (1942). *Insect Behavior.* University of Georgie, New York: Wiley-Interscience.

Merloti, P.E. (2004). Optimization Algorithms Inspired by Biological Ants and Swarm Behavior. San Diego State University, Artificial Intelligence Technical Report, CS550, San Diego.

Olugbemi, B. O. (2010). Influence of food on recruitment pattern in the termite, Microcerotermes fuscotibialis. *Journal of insect science (Online)*, *10*(154), (pp. 1-10). doi:10.1673/031.010.14114

PARC (2006). RMASE: Routing Modeling Application Simulation Environment. Available at: http://webs.cs.berkeley.edu/related.html

Perkins, C., & Royer, E. (1999). Ad-hoc on-demand distance vector routing. In *Proceedings of Second IEEE Workshop on Mobile Computing Systems and Applications* (pp. 90-100). doi:10.1109/MCSA.1999.749281

Ramos, V., & Almeida, F. (2000). Artificial Ant Colonies in Digital Image Habitats – A Mass Behavior Effect Study on Pattern Recognition. In: *Proceedings of ANTS'2000*, $2^{nd}$ *International Workshop on Ant Algorithms,* Brussels, Belgium (113-116).

Reinhard, J., & Kaib, M. (2001). Trail communication during foraging and recruitment in the subterranean termite Reticulitermes santonensis De Feytaud (Isoptera, Rhinotermitidae). *Journal of Insect Behavior*, 14(2), 157-171.

Resnick, M. (1994). Learning About Life. *Artificial Life*. **1**(1-3), (229-242).

Resnick, M. (1997). Turtles, Termites, and Traffic Jams: Explorations in Massively Parallel Microworlds. *Cambridge MA: MIT Press*.

Roth, M., & Wicker, S. (2003). Termite: ad-hoc networking with stigmergy. GLOBECOM '03, *IEEE Global Telecommunications Conference, 5, (*pp. 2937-2941).

Saleem, M., & Farooq, M. (2005). Beesensor: A bee-inspired power aware routing algorithms. In Proceedings EvoCOMNET, LNCS 3449 (pp. 136-146).

Saleem, M., Di Caro, G.A., & Farooq, M. (2010). Swarm intelligence based routing protocol for wireless sensor networks: Survey and future directions. *Information Sciences*, 181(20), 4597-4624. doi:10.1016/j.ins.2010.07.005

Semet, Y., O'Reilly, U., & Durand, F. (2004). An Interactive Artificial Ant Approach to Non-Photorealistic Rendering. *Springer-Verlag, K. Deb et al. (eds.): GECCO 2004, LNCS 3102,* (188-200).

Sztipanovits, J. (2004). Probabilistic wireless network simulator (Prowler), Retrieved from http://www.isis.vanderbilt.edu/Projects/nest/prowler/

Turner, J. S. (2010). Termites as models of swarm cognition. *Swarm Intelligence*, *5*(1), 19-43. doi:10.1007/s11721-010-0049-1

Wikipedia. (2003). *Simulation*, online Available from: http://en.wikipedia.org/wiki/Simulation.

Zhang Y. (2005). Routing Modeling Application Simulation Environment (RMASE), Available at: https://docs.google.com/file/d/0B-29IhEITY3bbGY2VVo2SGxxRFE/edit

Zhang, Y., Kuhn, L.D., & Fromherz, M.P.J. (2004). Improvements on Ant Routing for Sensor Networks. In M. Dorigo et al. (Eds.), ANTS 2004, LNCS 3172 (pp. 289-313).

Zhang, Y., Simon, G., & Balogh, G. (2006). High-Level Sensor Network Simulations for Routing Performance Evaluations. In: *Proceedings of* 3*rd International Conference on Networked Sensing Systems*, Chicago, 31 May-2 June 2006, pp. 1-4.

Zungeru, A.M., Ang, L.-M., & Seng, K.P. (2012a). Performance of Termite-hill Routing Algorithm on Sink Mobility in Wireless Sensor Networks. In Advances in Swarm Intelligence, Lecture Note in Computer Science, LNCS 7332(2), (334-343), Springer 2012.

Zungeru, A.M., Ang, L.-M., & Seng, K.P. (2012b). Classical and swarm intelligence based routing protocols for wireless sensor networks. Journal of Network and Computer Applications, 35(5), 1508–36.

Zungeru, A.M., Ang, L.-M., Prabaharan, S.R.S., & Seng, K.P. (2011). Ant Based Routing Protocol for Visual Sensors. *In A. Abd Manaf et al. (Eds.), ICIEIS 2011, CCIS 252 (pp. 250-264).*